# Controlled Evolution of Three-Dimensional Magnetic States in Strongly Coupled Cylindrical Nanowire Pairs

J Fullerton[1*], A Hierro-Rodriguez[2], C Donnelly[3], D Sanz-Hernández[4], L Skoric[5], D A MacLaren[1] and A Fernández-Pacheco[1,6**]

[1] SUPA, School of Physics and Astronomy, University of Glasgow, Glasgow, UK
[2] Depto. Fisica, Universidad de Oviedo, Oviedo, Spain
[3] Max Planck Institute for Chemical Physics of Solids, Dresden, Germany
[4] Unité Mixte de Physique, CNRS, Thales, Université Paris-Saclay, Paris, France
[5] Cavendish Laboratory, University of Cambridge, Cambridge, UK
[6] Instituto de Nanociencia y Materiales de Aragón, CSIC-Universidad de Zaragoza, Zaragoza, Spain

Email: *j.fullerton.1@research.gla.ac.uk; **amaliofp@unizar.es

**Abstract**
Cylindrical magnetic nanowires are promising systems for the development of three-dimensional spintronic devices. Here, we simulate the evolution of magnetic states during fabrication of strongly-coupled cylindrical nanowires with varying degrees of overlap. By varying the separation between wires, the relative strength of exchange and magnetostatic coupling can be tuned. Hence leading to the formation of six fundamental states as a function of both inter-wire separation and wire height. In particular, two complex three-dimensional magnetic states, a 3D Landau Pattern and a Helical Domain wall, are observed to emerge for intermediate overlap. The competition of magnetic interactions and the parallel growth scheme we follow (growing both wires at the same time) favours the formation of these anti-parallel metastable states. This works shows how the engineering of strongly coupled 3D nanostructures with competing interactions can be used to create complex spin textures.

## 1. Introduction

Magnetic nanowires have been widely studied due to their great potential for technologies including data storage, logic components and biological sensing [1-6]. Moreover, it has been shown that strong coupling between multiple nanowires can lead to novel emergent behaviours. For example, ultra-fast domain wall motion can be observed in synthetic anti-ferromagnetic nanowire systems [7] and coupled nanowire rings show promise for reservoir computing [8]. Previous research has mostly focused on rectangular nanostrips fabricated with thin film and planar lithography techniques. Now, with new fabrication methods, it is possible to investigate three-dimensional (3D) magnetic systems [9]. In their simplest form, these are cylindrical nanowires, where new types of magnetic states can be supported. In particular, both the cylindrical symmetry and the aspect ratio have been shown to have a significant effect on the magnetic states of the system [10-13]. Typically, for isolated nanowires with small aspect ratios, the magnetisation will lie in the radial plane. With increasing height, this is followed by a vortex with an axial core and finally a uniform axial state. In this work we will refer to these states as "Planar", "Vortex" and "Parallel" respectively. During fabrication, the aspect ratio will be constantly changing as the structure grows and therefore the magnetic state will continuously evolve through each of these states. Therefore, an important consideration is how a magnetic state may evolve during fabrication as this evolution can produce magnetic states that are distinct from the expected ground state.

Some of the fabrication techniques that allow this jump to 3D structures include: electrodeposition, two-photon lithography and focused electron beam induced deposition (FEBID). Utilising these techniques, the creation of more complex, 3D structures based on cylindrical nanowires can lead to the emergence of intricate domain structures [14-25]. One key area of interest for the development of spintronic devices is that of structures formed by strongly coupled nanowires [10, 26-34]. Examples of which include dense nanowire arrays, where promising spintronic properties may arise due to magnetostatic coupling; or if considering overlapped nanowires, exchange coupling is also introduced. In particular, FEBID has been shown to be an ideal nanofabrication technique for these investigations [22, 23, 26, 34]. For instance, strongly interacting helical nanowire pairs grown by FEBID can show locked pairs of domain walls in the case of separate wires and, when overlapped, chiral domain walls and topological 3D spin textures can form [26, 34]. However, these two previous works focused on helical nanowires, introducing a geometrically enforced chirality together with the magnetostatic and exchange energies. In this work, we focus instead on straight nanowires where geometric chirality does not play a role. Therefore, we can isolate the competition between exchange and magnetostatic energies in a coupled system. In these types of highly coupled structures, the magnetic state of one of the nanowires will have a large influence over the other, and vice versa, *from the very beginning of the fabrication process*. Here, through micromagnetic simulations, we investigate the evolution of magnetic states of highly coupled nanowire pairs grown in parallel as a function of inter-wire coupling and height. This procedure allows us to get an insight into how the magnetic states of the system varies with the strength of the inter-wire coupling, but also how a parallel growth strategy affects the final magnetic state of the system due to previously formed states.

## 2. Simulation Setup and Method

The micromagnetic simulation software Mumax3 [35] was used to mimic the growth of nanowires of varying separation. Cylindrical nanowire geometries with flat bottom and curved top caps were simulated to represent experimentally fabricated wires. A small mesh size of 1.25 nm was chosen to more accurately represent a cylindrical geometry and to be below the characteristic exchange length of the material. The diameters of the nanowires were set to d = 50 nm and the centre-to-centre inter-wire distance, s, was varied from 0 (fully overlapped) to 60 nm (fully separated): see figure 1a for these details. The Mumax3 in-built smoothing function was set to a value of 8 which, along with the small cell size, helped minimize staircase computational artefacts that can arise when simulating curved surfaces. We used micromagnetic parameters matching those typically obtained for FEBID cobalt [32]: saturation magnetisation, $M_s = 900 \times 10^3$ A/m (typically lower than the bulk); exchange stiffness, $A = 10^{-11}$ J/m, and zero magnetocrystalline anisotropy due to the nanocrystalline or amorphous nature of the material deposited in this manner [22, 34]). The geometry and material properties used here are also representative of common electrodeposited materials [32, 36-38].

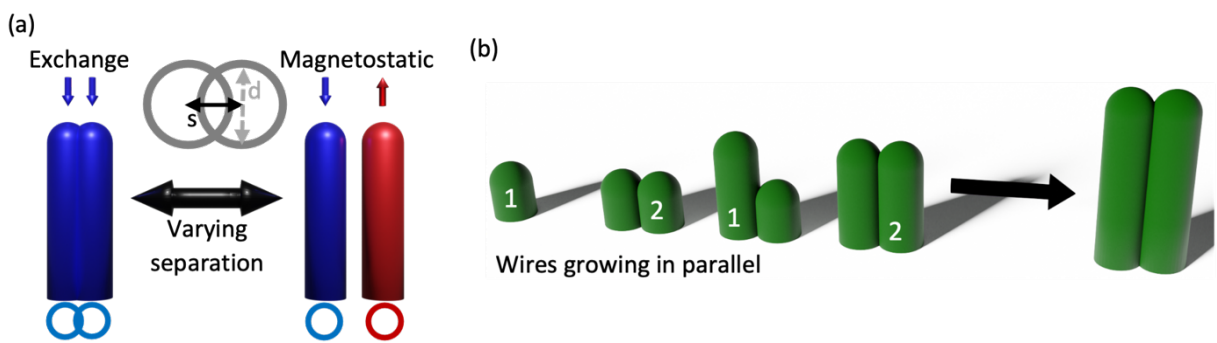

Figure 1: (a) Diagram showing how varying the separation of nanowires affects the magnetic interactions, with exchange dominating for small separation and magnetostatic dominating for large separations. The top inset shows that separation is defined as the distance between the centre of the two nanowires. Circles are shown below the wires to show how their cross-sections would appear. (b) Diagram showing parallel growth mechanism of nanowire pairs. Material is deposited sequentially between the nanowires until they reach the desired height.

To study the evolution of the states during growth, sections of 5 nm in height were added to the top of each of the nanowires sequentially. This parallel method of growth is commonly used in FEBID [26, 34] (shown in figure 1b). The simulated structures were computationally grown to a height of 240 nm, as a compromise between simulating structures with an aspect ratio high enough to strongly favour axial magnetic configurations, and a reasonable computational time (simulations would typically take 1-5 hours per growth step and were run on a NVIDIA GTX TITAN X graphical processing unit).

Each added section for every simulation step was given an initial random magnetic configuration. For all growth steps, the Landau-Lifshitz-Gilbert equation was integrated for the whole system (*i.e.*, including the spins in both new and old sections) while the maximum torque of the system was $> 10^{-4}$ T, with damping set at $\alpha = 0.5$. Once the above torque convergence criterion was reached, the magnetisation and energy densities were evaluated.

### 3. Results and Discussion
### 3.1 Evolution of magnetic states in nanowire pairs

During the simulated fabrication of the nanowire pairs, we expect that the magnetic state that forms at each growth step depends strongly upon the geometry of the system, *i.e.* how tall the nanowires have been grown and the separation between them. Figure 2a shows a vector representation of the six main states obtained from simulations for the ranges of separation (0 – 60 nm) and height (0 – 240 nm) under investigation. Four of these states are as discussed in section 1, appearing either for low heights or the extremes of separation: "Planar", "Vortex" and "Parallel"/"Anti-Parallel". The "Anti-Parallel" state occurs for separated nanowires where the uniform axial magnetic states form anti-parallel to each other due to the magnetostatic coupling (top right of figure 1a). All simulations show the general trend of planar –> vortex-like –> axial, as the height of the wires increase. Two less conventional states are observed for intermediate degrees of overlap. We denote these the "Landau pattern" and "Helical domain-wall" states. These two magnetic configurations feature net anti-parallel wires, but with a complex emergent 3D domain structure. The classification of these 6 magnetic states is based on the observation of sharp changes in the energy profiles at the transitions between them (further discussed in section S1 of the supplementary documentation). These transitions are used in figure 2b to generate a magnetic phase diagram as a function of height and separation. Figure 2c shows a map of the difference between exchange and magnetostatic energy density on the same axes, allowing the competition between energies to be evaluated for each state.

When the wires are fully overlapped (separation = 0 nm), they form a single cylindrical wire. During growth, the magnetic state evolves as previously reported in the literature for these geometries [10-13]: Planar – Vortex – Parallel. Vortex caps of the same chirality are observed at the top and bottom of the Parallel state to reduce surface charges.

For the next simulation (separation = 10 nm), a new state appears – the "Helical domain wall (DW) state" – evolving from the initial Planar and Vortex states (see figure 2b). This state is characterised by anti-parallel axial domains in the nanowires, with Néel caps at both ends and a DW in between them. In the bulk of the structure, a Bloch-type DW is favoured over a Néel-type DW to reduce magnetic volume charges. However, the Néel caps at the boundaries, along with the thickness of the overlapping region between the wires, also plays an important role in the DW structure, giving a modulated magnetic charged wall along the length of the system (details are contained in sections S2 and S3 of the supplementary information). As the Néel caps at both ends are antiparallel to each other, the spins forming the wall spiral continuously for 180° around the z-axis with height, forming a helix (a vector profile is shown in the examples of the helical DW state Figure 2a, more details about this state will be given in section 3.2). The name "Helical DW" is chosen from the profile of the spins in the region of overlap between the wires (this DW profile is also reminiscent of the "Twisted DW" seen in ref. 39). For this nanowire separation, and as the height increases during growth, the shape anisotropy of the long wire axis becomes increasingly dominant so that one of the axial domains is pushed out in favour of a Vortex state, finally followed by the

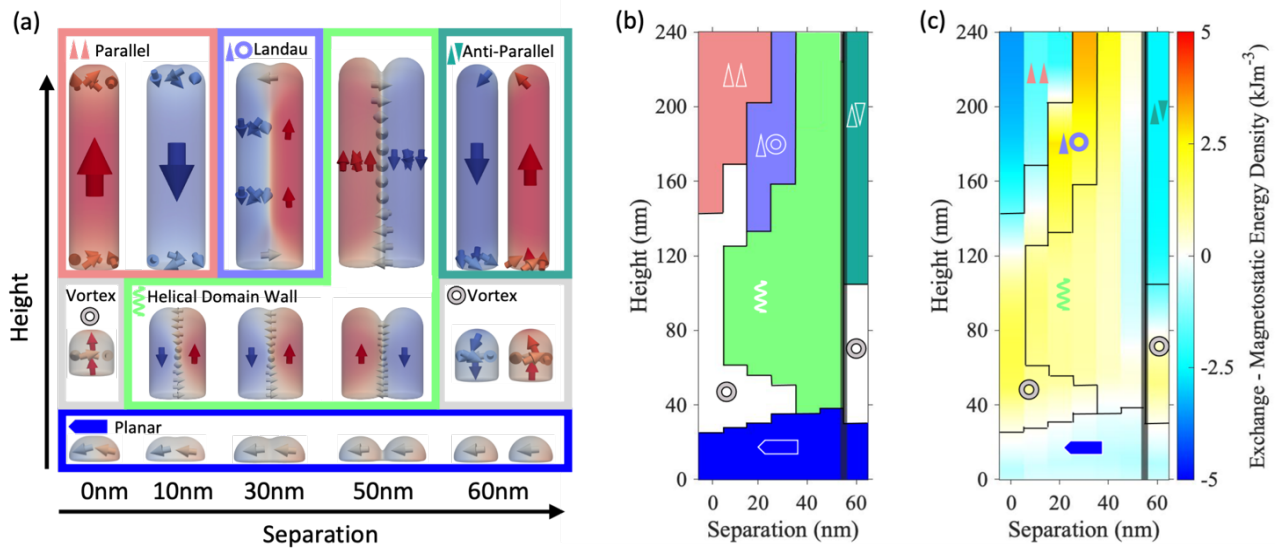

Figure 2: (a) A vector representation of the magnetic states that appear for varying separation and height. Each of the six states seen is given a colour and a symbol to match parts b and c of this figure as follows: Planar – blue pentagon pointing right; Vortex – grey donut; Helical domain wall – green helix; Parallel – Pink upward triangles; Landau – purple triangle and donut; and Anti-parallel – dark green opposite pointing triangles. (b) Phase diagram showing which of the 6 states are observed for each value of separation and height. Colours and symbols are defined in part a. (c) Map of the difference between exchange and magnetostatic energies for each value of separation and height, in order to determine how the two energies compete. For example, regions where exchange energy dominates are red/yellow in tone. Overlaid symbols and black lines are to define which magnetic state is present at each point on the map.

Parallel state. For the next value of separation (20 nm), the growth will begin in a similar fashion. However, now the system transitions from a Helical DW state to the Parallel state, not through a Vortex, but instead via what we denominate a 3D "Landau pattern" (named due to its resemblance to a Landau-Lifshitz pattern [40]). This state is similar to that previously described for rectangular magnetic systems, and consists of two anti-parallel domains [40, 41]. Here however, the magnetisation in one wire is strongly axial (red, net $+M_z$ domain in figure 2a), and takes the form of a vortex in the other (blue, net $-M_z$ domain). Due to its complex nature, this state will also be described in more detail in section 3.2.

For the subsequent simulations with larger separation between the wires, the final state of the structure (up to the full height that was simulated) consists of anti-parallel domains. The domain structure is defined by the thickness of the region of overlap between the wires. At a separation of 30 nm, the Landau pattern state is the final state achieved. For the next two values of separation (40 and 50 nm, figure 2b), the final state is the Helical DW. It can also be noted that at these separations, the Planar state transitions directly into the Helical DW state without going through a Vortex, due to the larger separation favouring anti-parallel magnetostatic coupling between the wires.

Finally, when the wires are fully separated (60 nm), the two individual wires follow a similar growth pattern to the fully overlapped simulations (separation of 0 nm, an isolated nanowire). However, for separated wires, the inter-wire magnetostatic coupling means that the vortex cores in each wire form anti-parallel to each other. Additionally, the formation of single domain axial states also occurs at lower heights than for the single wire, and the caps at the tops of the wires become reminiscent of Néel caps (due in part to their rounded geometry).

Figure 2c shows the difference between the exchange and magnetostatic energy densities, in order to evaluate which one dominates for each geometry. As expected, for colinear states (Planar or Parallel) the exchange energy is lower than the magnetostatic energy. Conversely, more complex magnetic states (Vortex, Helical DW, Landau pattern) correspond to a higher exchange energy density, due to the formation of non-collinear spin configurations. Out of all the states found in the simulations, the Landau pattern is the most costly in exchange energy density, followed by the Helical DW and Vortex states, which have similar values. These results show that for the design of coupled geometries capable of stabilising complex spin textures, care must be taken to ensure that the coupling between the structures has an appropriate intermediate value to allow the magnetic energies to compete with each other (*i.e.*, not allowing one to dominate over the other). In the case of this study, this is observed when the overlapped wires form a net magnetic anti-parallel alignment to each other.

### 3.2 Emerging three dimensional magnetic states for intermediate overlapping

For intermediate separation, the system tends to evolve towards states where the magnetostatic energy is minimized by aligning the overall magnetisation of both wires anti-parallel to each other. However, in doing so, an increase in exchange energy takes place due to the formation of complex 3D domain structures as a consequence of this anti-parallel alignment. The type of configuration that forms is defined by the thickness of the overlapping region between the wires. A comparison between the two states emerging for intermediate overlapping is shown in Figure 3 with views along each axis. The viewing directions for figures 3d-f are shown in 3a-c respectively. Figure 3d shows a side view for both states on the left and right: a colour plot of the $M_x$ component on a slice in the overlapping region of the wires and an isosurface of spins within 10% of the -$M_y$ component to visualise the domain wall. The top central inset of figure 3d corresponds to the evolution of $M_x$ as a function of height along the Z-axis for the middle overlapping area, as marked by coloured dashed lines overlaid on the simulation plots. Whereas the bottom inset shows a detailed zoom-in of the magnetic configuration for the Landau pattern where the DW meets the surface. Figures 3e and 3f complement figure 3d by plotting a front view of the $M_z$ isosurfaces (figure 3e) and XY projections of the magnetisation vector coloured by $M_z$ (figure 3f) for seven representative heights along the z axis (locations of which are marked with grey dashed lines in figure 3e).

We first focus on the Landau pattern state. In this case, the overlapping region is nearly as thick as the wires themselves (in the y-direction, shown in figure 3d), and thus allows for the extension of the Néel caps along the surface of the structure (shown as yellow on the left and purple on the right in figure 3d). This magnetic state reduces surface charges and creates an S-shaped DW (grey isosurfaces in figure 3d) between the wires that only has one point on each of its edges where it meets the surface of the structure (see left panel in figure 3d and the corresponding bottom inset) [40, 41]. A notable feature of this magnetic state is that the anti-parallel domains are not equivalent: whereas in one wire (the net $+M_z$ domain, red in figure 3e) the magnetisation is strongly aligned with the z-axis, the domain in the other wire (the net $-M_z$ domain, blue in figure 3e) takes the form of a vortex. In fact, this structure contains one single vortex tube that begins and ends at the two points where the Landau domain meets the surface of the structure. However, the vortex does not follow the Landau domain within the bulk of the structure (in-between cross-sections iii and iv in figure 3e). Instead, it extends through the net $-M_z$ domain, resulting in a vortex extending along the length of this wire (distinguishing this state from a Landau-Lifshitz pattern in a rectangular geometry [40]). This is due to the vortex being able to adapt easily to the cylindrical geometry of one of the wires. Cross-section iv in figure 3f highlights this vortex-axial configuration in the net structure of anti-parallel domains. The other cross-sections in this figure show the transition between the ends, where Néel caps are formed at top and bottom, and the body of the structure, where the magnetisation evolves towards a vortex-axial configuration. The Landau state can be distinguished from others, including the Helical DW state discussed below, by monitoring the energy density of the simulation during growth. When the Landau pattern forms, the magnetostatic energy reaches a plateau (as the surface charges produced by the Landau domain are consistent regardless of the growth step), while exchange energy continues to rise (see section S1 in the supplementary documentation).

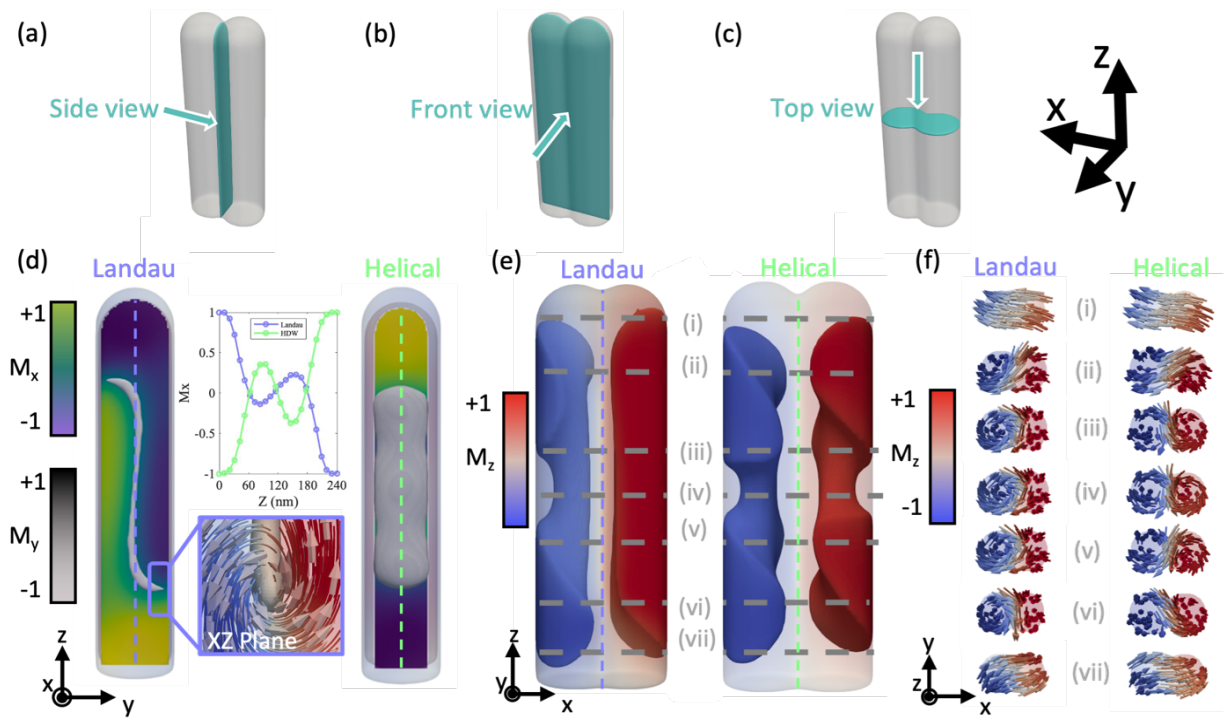

Figure 3: (a, b, c) Viewpoints for figures d, e and f respectively. (d) Left and right sections show a side view of the Landau pattern and helical DW state respectively, coloured by the $M_x$-component in a central slice. A 2% isosurface in each of spins within 2% of the negative y-axis (grey surface) is shown to highlight the different domain structures between the simulations. The top central inset shows a plot of the $M_x$ component vs Z-coordinate in the centre of the structure (purple and green dotted lines respectively) to highlight the modulation seen in the profile of the domain wall. The bottom inset shows a vortex in the XZ plane around the location where the Landau domain meets the surface of the structure. (e) Left and right sections show a front view of 10% +/- $M_z$ isosurfaces of the Landau pattern and helical DW state, respectively, to highlight the differences in the anti-parallel domains of both states. The Landau pattern shows a vortex-axial configuration while the helical DW shows a double vortex state. Coloured dotted lines show the location of the spins plotted in the top inset of part d and grey dotted lines are the locations of the cross-sections shown in f. (f) Left and right sections show XY cross-sections of the Landau pattern and helical DW state respectively, coloured by the $M_z$ component at the locations of the grey dotted lines in part e. These are to show the differences in the anti-parallel domain structure between the two states, and also the transition between the caps and the bulk of the structures.

The second emerging 3D magnetic state observed in the simulations for intermediate overlapping is the Helical DW. Here, the region in-between the wires is smaller than in the previous case discussed (30 nm separation), inhibiting the formation of a Landau pattern due to the high cost in exchange energy. Instead, here only a DW is formed between the wires, as briefly explained in the previous section. In comparison with the Landau pattern, the two anti-parallel domains of the wires in any Helical DW state are equivalent to each other. When the state first forms, the anti-parallel domains are strongly axial. However with increasing height, they become more vortex-like (with opposite cores and circulations, and hence, the same chirality). Cross-section iv in figure 3f shows this double vortex configuration of the domains in the mid-point of the structure.

Interestingly, when plotting the $M_x$ component across the Néel caps and DW in both states (top inset in figure 3d), a modulation in the profile of the DW is observed for fully grown states (see section S2 of the supplementary documentation for further discussion). This modulation forms to oppose magnetic charges created at the Néel caps at the top and bottom of the structure and is discussed further in section S3 in the supplementary information.

The emergence of both the Helical DW and the Landau pattern show the influence that competing magnetic coupling between wires can have on the magnetic states available to the system. However, as discussed in the introduction, we also want to investigate the effect of the growth history of the system. Therefore, in the next section we examine the impact of the parallel growth mechanism on the states formed.

### 3.3 Induction of metastable states via the parallel growth method

Given that we are simulating a parallel growth mechanism, (*i.e.*, two wires growing at the same time in small sequential steps) the equilibrium magnetic state reached may not necessarily be the ground state. To investigate the effect of the growth mechanism, the regions of stability for each of the fully grown overlapped states (*i.e.*, Parallel, Landau pattern and Helical DW) were investigated between separations of 10 and 50 nm (as these were the values of separation that marked out the limits of structures consisting of two overlapped nanowires). The output of a growth simulation was selected for each state: s = 10 nm for Parallel, 30 nm for Landau pattern and 50 nm for Helical DW. The separation of the wires was then increased or reduced by one cell at a time. Any new cells introduced were given a random magnetisation. At this new value of separation, the system was allowed to evolve dynamically to ascertain if the state in question would be stable (to the same criterion as the growth simulations described in section 2). If the state remained stable, the energy was evaluated and the separation was altered by one cell again. This process was repeated until either the state was no longer stable or the full range of separation had been probed.

Figure 4 shows the energy of each of these states at every value of separation where they remained stable. From this protocol, it is observed that the Parallel state remains metastable for all degrees of overlap, despite not forming during our simulated growth procedure for separation $\geq$ 30 nm. The energy density of the Parallel state increases with separation due to magnetostatic energy becoming more costly (this relationship varies from a linear trend for separations above 45 nm, due to subtle changes in the magnetic configuration of the caps of the structure). The Helical DW state however, shows the opposite trend, with energy density increasing with decreasing separation until it collapses into the Landau pattern below separations of 31.25 nm. The trends for these two states show the expected trade-off between the magnetostatic and exchange energies, with a cross-over in ground state as magnetostatic energy dominates at higher separations and exchange dominates at lower separations (figure 1a). The Landau pattern shows a similar energy trend to that of the Helical DW, and is stable between separations of 22.5 and 38.75 nm. Below 22.5 nm, the Parallel state is the only stable configuration of the system, and above 38.75 nm the overlapping region between the wires becomes too small and prevents the Landau pattern from forming (as discussed in section 3.2).

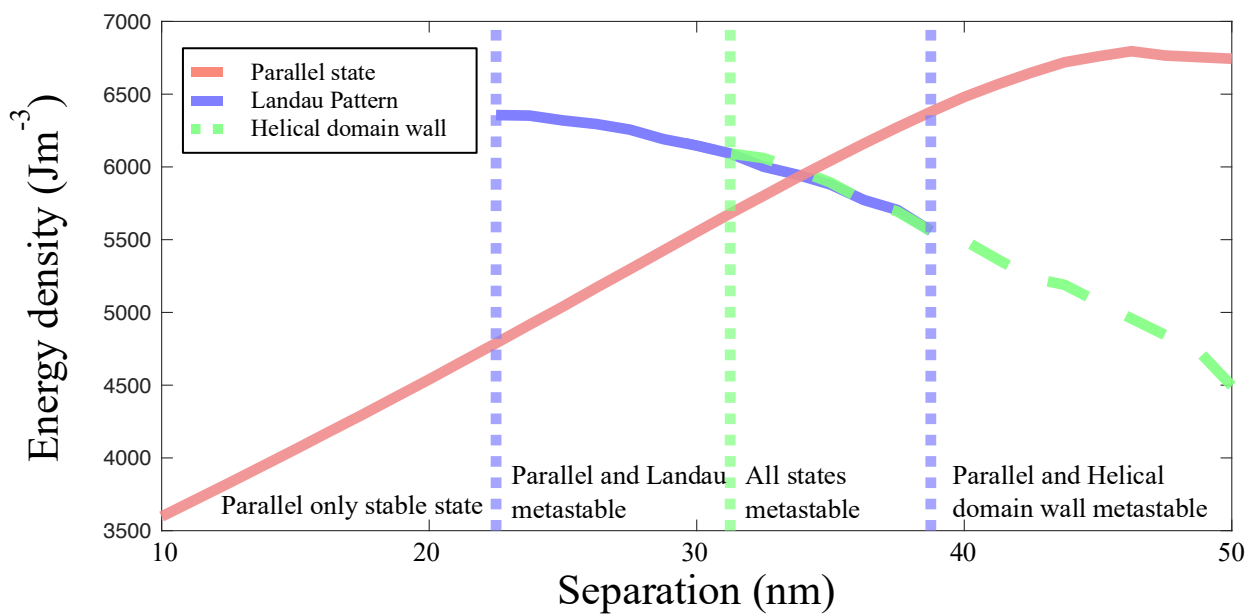

Figure 4: Relative energies of metastable states in fully grown (240 nm) nanowire pairs for different inter-wire separation; Parallel – red, Landau pattern – purple and Helical DW – green. The energy profile of the Helical DW is shown using a dashed line so that both remain visible when they overlap The energy of each magnetic state was evaluated for each value of separation where it was observed to be stable. The regions of stability of the Landau pattern (intermediate separation values) and Helical DW (largest separation values) are marked out with vertical dotted lines of their corresponding colour. The parallel state is always stable for all values of separation.

As shown in figure 4, there are three regions where multiple states coexist. Above 38.75 nm the Parallel and Helical DW states can form. However, the saving in magnetostatic energy from having an anti-parallel alignment between the wires means that the Helical DW is the ground state. This fact lines up with the growth simulations, as the Helical DW was achieved as the final state for separations of 40 and 50 nm. Between 31.35 and 38.75 nm, all three states are metastable with each other. The energy densities of the Landau pattern and Helical DW have very similar values in this region as their respective savings of magnetostatic and exchange energy over each other effectively cancel out in this region. The ground state switches in this region from an Anti-parallel alignment of wires to a Parallel one below 34 nm, this remains the case for any nanowire pair with a smaller value of separation.

The most revealing aspect of this method of simulation is highlighted in the final region of metastability. Between 22.5 and 31.25 nm only the Landau pattern and the Parallel state can exist, but the most energetically favourable state is the Parallel state, due to its savings in exchange energy. Nevertheless, for separations of 30 nm, it is the Landau pattern that forms during our growth simulations. In the parallel pair growth scheme utilised here, the small aspect ratio seen at the start of growth favours an anti-parallel alignment between the wires. As the aspect ratio increases, the parallel state will become more energetically favourable [10-13]. However, an energy barrier may form, preventing the annihilation of one of the domains. For lower separations, this energy barrier can be overcome (*e.g.*, for a separation of 20 nm, figure 2b). Conversely, for a separation of 30 nm, the energy barrier is high enough to prevent this transition. A common method to explore the magnetic state of a nanowire involves relaxing the system from a saturated state [42, 43]. In this case however, due to the metastability of the Parallel state at all separations, this method would not return the overlapped anti-parallel states presented here. Consequently, we would be building an inaccurate picture of the effect of the magnetic coupling. Whereas the controlled-evolution method used here gives a more comprehensive depiction of this type of system.

The results presented here highlight the influence and control that the fabrication method can exert on the magnetic state of the system. In particular, a direct-write technique such as FEBID allows for significantly different growth strategies to be employed to realize the same geometry (*e.g.,* fully growing one wire before starting on the second may produce a different as-grown magnetic remanent state).

## 4. Conclusion

We have presented a simulation study of the control of magnetic states through both the tuning of coupling between nanowire pairs and parallel fabrication of the nanowires. Together, this leads to the formation of a wide variety of 3D magnetic states. Included in these are the Landau pattern and modulated helical DW, which both only form for limited combinations of height and separation. The impact that the fabrication process can have on the magnetic state of the system is revealed. The parallel-growth method followed here favours an anti-parallel magnetic coupling between the wires, even when not the ground state, which leads to complex 3D domain structures forming between the two opposite domains in both wires. This result exemplifies the influence that the fabrication method has over the magnetic configuration of a given system.

The growth simulation method presented here could be expanded to mimic different fabrication techniques (*e.g.*, electrodeposition, electroless deposition and atomic layer deposition) [44, 45]. It would also allow for the addition of further complexities. For example, a radial growth process could be employed in the place of, or together with the axial process shown here [46, 47]. Magnetocrystalline or magnetoelastic anisotropies could be introduced, thereby adding a third magnetic energy to the competition [48]. Furthermore, we could consider variations of composition and geometry along the length of the wire [15, 16, 49,50]. The understanding how different degrees of magnetic coupling may affect magnetic states in nanostructures is important for many sub-areas of nanomagnetism, from biomagnetism to spintronics, and will assist in the understanding and design of coupled 3D nanostructures.

**Acknowledgments**

The authors would like to thank Naemi Leo and Jakub Jurckyck from CSIC in Zaragoza and Miguel A. Cascales Sandoval from the University of Glasgow for fruitful discussions. We would also like to thank Sam McFadzean at the Kelvin Nanofabrication centre in Glasgow for technical support, and the University of Glasgow where this work was carried out. This work was supported by the EPSRC and the Centre for Doctoral Training (CDT) in Photonic Integration and Advanced Data Storage (PIADS), RCUK Grant No. EP/L015323/1, the European Community under the Horizon 2020 Program, Contract No. 101001290 (3DNANOMAG), the MCIN with funding from European Union NextGenerationEU (PRTR-C17.I1), and the Aragon Government through the Project Q-MAD. A.H.-R. acknowledges the support from European Union's Horizon 2020 research and innovation program under Marie Skłodowska-Curie grant ref. H2020-MSCA-IF-2016-746958, from the Spanish MICIN under grant PID2019-104604RB/AEI/10.13039/501100011033 and from the Asturias FICYT under grant AYUD/2021/51185 with the support of FEDER funds. *CINN (CSIC – Universidad de Oviedo), El Entrego, Spain* L.S. acknowledges support from the University of Cambridge (EPSRC Cambridge NanoDTC EP/L015978/1) D.H. acknowledges Unité Mixte de Physique, CNRS, Thales, Université Paris-Saclay, 91767, Palaiseau, France. C.D. acknowledges funding from the Max Planck Society Lise Meitner Excellence Program.